# AN IMPROVED ALGORITHM FOR RECONSTRUCTING SINGULAR CONNECTION IN THE MULTI-BLOCK CFD APPLICATIONS

WANG Yong-Xian（王勇献）[1,2,*], ZHANG Li-Lun（张理论）[1,2], CHE Yong-Gang（车永刚）[1,2], XU Chuan-Fu（徐传福）[1], LIU Wei（刘巍）[1], LIU Hua-Yong（刘化勇）[3], WANG Zheng-Hua（王正华）[1]

[1] College of Computer, National University of Defense Technology, Changsha 410073, China
[2] Science and Technology on Parallel and Distributed Processing Laboratory, National University of Defense Technology, Changsha 410073, China
[3] State Key Laboratory of Aerodynamics, Mianyang 621000, China

**Abstract:** In this paper, an improved algorithm is proposed for the reconstruction of singularity connectivity from the available pairwise connections during preprocessing phase. To evaluate the performance of our algorithm, an in-house CFD code, in which high-order finite-difference method for spatial discretization, running on the Tianhe-1A supercomputer is employed. Test cases with a varied amount of mesh points are chosen, and the test results indicate that the improved singular connection reconstruction algorithm can achieve a 2000× speedup at least compared with the naive search method adopt in the former version of our code. Moreover, the parallel efficiency can be benefited from the strategy of local communication based on the new algorithm.

## 1  Introduction

Most techniques for doing computational fluid dynamics (CFD) rely on the subdivision of physical space into a grid of discrete grid points of computational space in which values of flow variables can be defined, followed by transformation of the governing equations (differential form or integral form) into algebra equations using finite difference method, finite volume method or finite element method. The simulation results from the final converged state by advancing a starting solution through a sequence of iteration steps. Multi-block structured grid has been widely used in the CFD simulations for its inherent advantages of easy-manipulated, simple-implemented, accurate calculate ability and strong boundary-dealing ability, especially in the case dealing with aircraft flow fields formed of complex shapes. There are three types of connections between each pair of neighboring blocks in a typical multi-block structured grid, i.e. the 1-to-1 one, patched one and overset one, illustrated in Fig. 1. In this paper, we mainly focus on the grid with 1-to-1 connection type and ignore the other two types.

Foundation items: Supported by the National Natural Science Foundation of China (61379056, 11272352); the Open Research Program of China State Key Laboratory of Aerodynamics (SKLA20130105).
Corresponding author: Wang Yongxian, Associate Professor, E-mail:yxwang@nudt.edu.cn

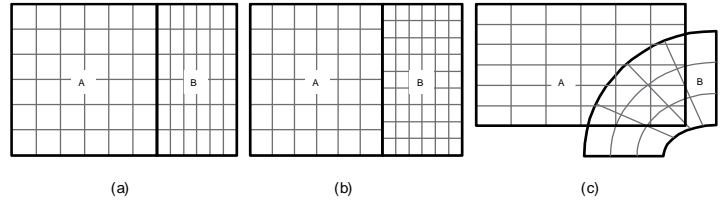

Fig. 1 Three types of connections in the structured grid. (a) 1-to1 type (b) patched type (c) overset type

In the three-dimensional 1-to-1 type multi-block grid, each pair of neighboring blocks share a common interface (face or portion of a face) and they are point-matched along that interface (i.e. all the boundary points of neighboring blocks coincide). We define the face in the common interface as a connection face, or a connection edge, connection node when the interface degenerate as an edge, a node respectively. A connection node surrounded by more than two blocks is a geometrical singularity node, or singularity node for short. It can further be classified into either a physical boundary singularity node, if the singularity node is located on a physical boundary of flow-field, or an internal singularity node otherwise. We call the connection edge (face) singularity edge (face) when it is composed of singularity nodes only. Finally, any connectivity across common singularity faces (along with singularity edges and singularity nodes) is generally called singularity connectivity.

In the CFD simulations, applying the finite difference discrete method or finite volume discrete method, the flow variables can be stored at either the centroids of the grid cells (cell-centered scheme) or the grid points (cell-vertex scheme). When adopting cell-vertex scheme in CFD applications with the multi-block grid with singularity connectivity, the flow variables at the grid points may create multiple copies, and each copy is corresponding to one of the blocks surrounding the singularity point. For example, in the two dimensional multi-block structured grid, illustrated in Fig 2(a), a singularity node $P$ (shown by thick dot) is shared by three surrounding blocks, via block 1, 2 and 3. As a result, there will be $m$ ghost nodes (here $m=3$), $P_1, P_2, \cdots, P_m$, distributed in block 1, 2, …, $m$, respectively. During the time advancing phase of multi-block structured grid CFD simulations, a large scale of linear systems can be formed in each single block, and the solution at each grid point represents the flow variables in current time step. These linear systems are solved independently, which may lead to different solutions (also named flow state variables) $Q_1, Q_2, \cdots, Q_m$ in the ghost nodes $P_1, P_2, \cdots, P_m$, although they are corresponding to a same grid point $P$ in the flow domain. In order to void misleading result, a correction step is introduced. A new quantity $\overline{Q}$ is constructed from $Q_1, Q_2, \cdots, Q_m$ by letting $\overline{Q} = \varphi(Q_1, Q_2, \cdots, Q_m)$ once original flow variables on the ghost nodes are obtained before advancing to the next time step. This additional correction step guarantees that all ghost nodes located in different blocks, corresponding to a same singularity node, will have the same new state value. To achieve this, the topological relations, or the connectivity, among different blocks, must be built and stored in the pre-process phase.

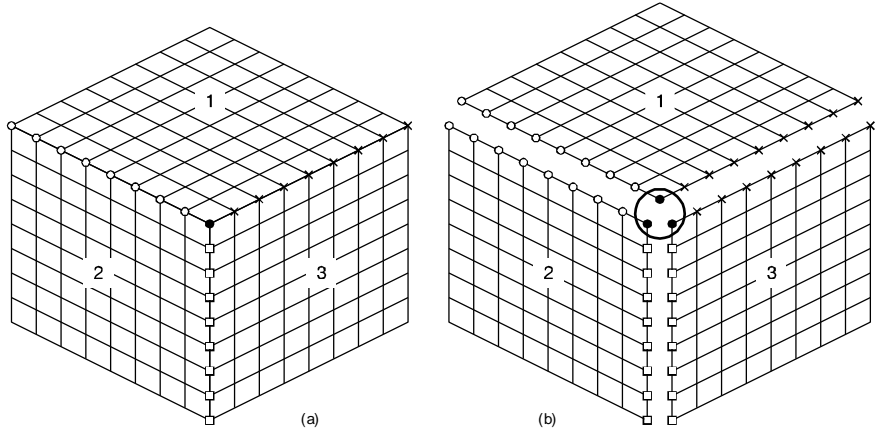

Fig. 2 Singular node in the 1-to-1 multi-block grid. (a) A singular node shared by three neighboring blocks
(b) Three nodes as the copies of the original singular node

Apparently the best time to build and record these topological relations is the grid generation phase, however the prevailing tools of grid generation can only export the connectivity between each pair of blocks, and the information of singularity nodes/edges is completely lost. How to efficiently reconstruct the connectivity information among multiple blocks based on the available pairwise connectivity, especially finding the correspondence of each singularity point and its ghost nodes distributed multiple blocks, will play an important role in the CFD simulation [1-3]. For the parallel CFD simulation application with huge cell or grid point size, it remains a large amount of blocks in the grid in an attempt to get high parallelism degree, as a result, the performance of reconstructing singularity connectivity will become another significant issue.

In this paper, we propose a fast reconstructing singularity connectivity algorithm for the multi-block structured grid, and discuss the performance of the new algorithm. The remainder of this article is organized as follows. The description of the problem is given in Sect. 2, followed by two reconstructing singularity algorithms, the original algorithm 1 and improved algorithm 2, and complexity analysis. In Sect. 3 a couple of numerical experiments are tested, along with the results being evaluated and discussed. Finally Sect. 4 concludes our work.

## 2 Reconstructing singularity algorithms

### 2.1 Problem Description

Considering a typical structured grid CFD simulation, the three-dimensional simulation domain (in the physical space) is expected to be discretized first, in an attempt to form a discretizing grid points set, $G'$, containing $V'$ points. The discretizing grid points set is then partitioned into $n$ blocks. Each block contains a portion of grid points, using domain decomposition methods in accordance with requirements of flow field shape and simulation calculation. As a result, the grid points located on the common surface of a pair of neighboring blocks will have two copies, one in the left block, and the other in the right block. Take the *l*-th block for an example, suppose the local grid points are organized in a hexahedral structure with size of $V^{(l)} = N_i^{(l)} \times N_j^{(l)} \times N_k^{(l)}$. Thus each local grid point in the *l*-th block has a (global) computational coordinate $(l;i,j,k)$, where $l = 1,2,\cdots,n$ is the block number, and $(i,j,k)$ is the local computational coordinate ($i = 1,2,\cdots,N_i^{(l)}$; $j = 1,2,\cdots,N_j^{(l)}$; $k = 1,2,\cdots N_k^{(l)}$). As the grid points in physical space have their own computational coordinates, for distinguishing the points in physical space and the ones in computational space, we use the term "grid nodes" instead of "grid points" to denote the points in the computational space.

In the computational space, let $G_l$ be the set of all grid nodes in the *l*-th block, i.e.:

$$G_l = \{(l;i,j,k) : i = 1,2,\cdots,N_i^{(l)}; j = 1,2,\cdots,N_j^{(l)}; k = 1,2,\cdots N_k^{(l)}\}$$

Let $G = \bigcup_{l=1}^{n} G_l$ be the set of grid nodes in whole grid. We denote $V^{(l)} = |G_l|$ and $V = |G|$. Note that the points number in the physical space *V'* is not larger than the one in the computational space, i.e. $V' \leq V = \sum_{l=1}^{n} V^{(l)}$, due to the existence of node copies mentioned above. The equivalence relations between grid nodes in the sense of mathematics can be introduced.

**Definition 1[Equivalence]**: In the computational space of multi-block structured grid $G$, a pair of grid nodes $p = (l;i,j,k)$ and $p' = (l';i',j',k')$ are equivalent, donated as $p \sim p'$, if they both correspond to the same grid point in the physical space $q \in G'$.

The equivalence relation defined here satisfy the properties of reflexivity, symmetry and transitivity mathematically. On the basis of equivalence relation, we can further define the equivalence class of a grid node and the quotient set of the grid.

**Definition 2[Equivalence class]**: Suppose $p \in G$ is a grid node in the multi-block structured grid $G$. The node set $[p] = \{p' \in G : p' \sim p\}$ is the equivalence class of node p if including all nodes equivalent to *p* in grid $G$.

**Definition 3[Quotient Set]**: The set of all equivalence classes (given an equivalence relation ~) in $G$ is denoted by X/~, and is defined as the quotient set of $G$ by ~.

We are ready to define the singularity connectivity in the multi-block structured grid based on the equivalence class and quotient set.

**Definition 4[Singularity point/node]**: For a given equivalence class $P \in G/\sim$ in a multi-block structured grid *G*, with $|P| > 2$, all elements (nodes) in P correspond to a same grid point $q \in G'$ in physical space. The special grid point *q* is known as a singularity point. All nodes in *P* are called singularity nodes, or the copy nodes of *q*.

Now we can give a formal description about reconstructing singularity connectivity problem: for a given multi-block structured grid *G*, our target is to find out all singularity points (nodes) in line with the pairwise connectivity between each pair of neighboring blocks in it. There is no need for CFD applications to provide each and every singularity point in the physical space in practice, instead, only equivalence classes are needed.

### 2.2 Reconstruction algorithm and complexity analysis

From the previous description, we are clear that the essence of reconstructing singularity connectivity for the grid *G* is to provide the quotient set of equivalence relation based on pairwise connectivity in the grid *G*. We can easily have the original version of reconstruction algorithm listed in Algorithm 1 by the transitivity property of the equivalence relation.

---

Algorithm 1: Original algorithm for reconstructing singularity

INPUT: grid *G*, and the equivalence class derived from pairwise connectivity $E = \{(p, p'), p \in G, p' \in G\}$

OUTPUT: set *D* of equivalence classes with more than 2 elements.

Steps: 
1. Construct the candidate node set $C = \{p \in G:$ there exists $p' \in G$, such that $(p, p') \in E$ or $(p', p) \in E\}$
2. Initialize set of equivalence classes D to be empty
3. For each candidate node $p \in C$, do step 3.1 ~ 3.4
   - 3.1 IF (*p* is not contained in any element set of *D*) THEN add the singleton {*p*} to *D*
   - 3.2 Find the unique element *P* in *D*, such that $p \in P$
   - 3.3 Find the element set *P'* equivalent to node *p* in E, and add it to *P*, i.e.: $P \leftarrow P \cup P'$
   - 3.4 Remove the element set in *D* with the number of elements equal 2

---

In implementation of Algorithm 1 by modern programming language, if a set containing grid nodes is expressed by a array, a common data structure, and let the number of candidate nodes $N = |C|$, the maximal number of nodes contained in set *D* is $|C| = O(N)$ during the whole processing phase. The complexity of step 1 in Algorithm 1 is $O(N)$, so is in step 3.1 and step 3.4. By linear time searching algorithm, the computational complexity of both step 3.2 and step 3.3 is $O(N^2)$. So the total time complexity of Algorithm 1 is $O(N^2)$.

Algorithm 1 was employed in the original version of our in-house CFD code with the finite difference method coupled with cell-vertex scheme. Its main advantage is simple and easy for programming, while its computational complexity is far from satisfactory. For a test case with a structured grid composed of 506 blocks and $N$=17.9M grid nodes, the complexity, $O(N^2)$, will reach $8.28\times10^{12}$, which will cost 354 seconds in a machine with single 2.93GHz CPU processor. With the increasing total number of grid nodes and the number of blocks due to the parallel computation, the huger $N$ will make the running time intolerable. To improve Algorithm 1 by optimizing the data structure and key steps is necessary. The improved algorithm is listed in Algorithm 2.

---

Algorithm 2: Improved algorithm for reconstructing singular singularity

INPUT: grid $G$, and the equivalence class derived from pairwise connectivity $E = \{ (p, p'), p \in G, p' \in G \}$

OUTPUT: set $D$ of equivalence classes with more than 2 elements.

Steps:  1. Construct the set of candidate grid nodes

   1.1 Construct the set of nodes on the block edges H={ $p \in G$: p is on an edge }

   1.2 Construct candidate nodes set $C = \{ p \in G$: there exists $p' \in G$, such that $(p, p') \in E$ or $(p', p) \in E\}$

   1.3 Construct the candidate edges set $F = H \cap C$

   1.4 For each pair of equivalent nodes $p$ and $p'$ in E, do step 1.5

   1.5  IF (either $p$ or $p'$ is in $F$) THEN add the other one to $F$ to ensure $F$ contains both $p$ and $p'$

 2. Construct the candidate set of equivalence classes

   2.1 Initialize set of equivalence classes to be $D=\{ \{p\}: p \in F \}$

   2.2 For each pair of equivalent nodes p and $p'$ in $E$, do step 2.3

   2.3  IF ($p \in F$ and $p' \in F$) THEN merge the subset of $D$ contains p and the subset of $D$ contains $p'$

 3. Remove the element set in $D$ with the number of elements equal 2

---

As most candidate nodes in set $C$ are located at the inner of block surface in Algorithm 1, and their equivalent class only contains two elements. As a result these candidate nodes will be removed at step 7 in Algorithm 1. We start our work from initial candidate set of nodes $F$ (see step 1.3) with smaller size in Algorithm 2, and the grid nodes contained in $F$ are expected to be on the edge of a block. Further analysis indicates that, for any singularity points, there exists at least one equivalent node located on the edge of a block. So a reasonable strategy is to collect all nodes on edge of blocks to form an initial candidate "seeds" set, and in step 1.4-1.5 of algorithm 2 we try to find those "missing nodes", which are not on any edge of blocks but are still equivalent to a node in $F$, and add them back to $F$.

Let $M = |H|$ be the total number of nodes located on the edges in the grid blocks, and $N = |C|$ be the number of nodes located on the faces in the grid blocks as defined before, and in most cases, it has $M << N$. To improve the running efficiency of the new algorithm, we redesign the data structure and take some extra measures as follows. (1) A binary sort tree of grid node is employed in constructing the candidate set F of nodes, which makes the time complexity approximate $O(\log |F|) = O(\log M)$ for searching, or inserting an element in set $F$. (2) To speed up the searching grid nodes equivalent to $p$ in set $E$, $E$ is repartitioned into $n$ subsets, namely buckets, corresponding a single block of grid each in the preprocessing. The size of elements to be searched will decrease by a factor $1/n$, leading to a $O(N/n)$ time cost. (3) A two-way circular linked list as the container of nodes is applied to construct the equivalence classes in a constant time period.

After adopting all these improvement, the computational time complexity of each step in algorithm 2 is listed as Table 1, making the total time complexity to be $O(N \log M)$.

Table 1　time complexity of algorithm 2

| step | time complexity |
|---|---|
| step 1.1~1.5: | $O(N \log M)$ |
| 　step 1.1 | 　$O(M)$ |
| 　step 1.2 | 　$O(N)$ |
| 　step 1.3 | 　$O(M \log M)$ |
| 　step 1.5 | 　$O(N \log M)$ |
| step 2.1~2.3: | $O(N \log M)$ |
| 　step 2.1 | 　$O(M)$ |
| 　step 2.3 | 　$O(N \log M)$ |
| step 3: | $O(M)$ |

## 3　Results and discussion

### 3.1　Test platform and results

To evaluate the performance of our reconstructing singularity connectivity algorithms, we conduct several numerical experiments on TianHe-1A high-performance computing system. Considering the reconstruction of singularity connectivity routinely acts as a phase of preprocess for CFD simulations, our tests only run in a serial mode on a single machine node with Intel Xeon X5670 6-core 2.93GHz CPU, with memory size of 48GB. In order to remain the consistent with the CFD solver code, the reconstructing singularity connectivity code is programmed using Fortran 90 language and compiled by the Intel Fortran (version 11.1) with `-O3` compiling option.

The test cases include flow field simulation for 14 different configurations of four types of aircraft shapes (Table 2). The cost of wall time for the singularity connectivity reconstruction is measured with 14 different configurations respectively in original algorithm 1 ("old algorithm") and improved algorithm 2 ("new algorithm"), measured by the Fortran subroutine `SYSTEM_CLOCK()`. Each running time reported is the least one during 5 repetition runs. The last column of table 1 shows the speedup of new Algorithm 2 compared to the old one (Algorithm 1). It costs too much time for Algorithm 1 to run the last two cases (we terminate the running after 3 days), so we mainly take the remaining 12 cases into consideration.

Table 1 Running time of reconstructing the singular connection for test cases with 14 different configurations

| Mesh ID | # mesh cells | #blocks (n) | #singular points ($N_s$) | Running time (sec.) | | Speedup |
|---|---|---|---|---|---|---|
| | | | | Alg. 1 | Alg. 2 | |
| DLR-F6 | 16 513 024 | 355 | 32270 | 142.667 | 0.044 | 3242 |
| | | 381 | 35409 | 168.643 | 0.050 | 3400 |
| | | 472 | 46175 | 265.941 | 0.078 | 3392 |
| airfoil NACA0012 | 13 104 000 | 45 | 7071 | 16.915 | 0.006 | 3020 |
| | | 64 | 13396 | 27.379 | 0.012 | 2281 |
| | | 168 | 25108 | 77.662 | 0.027 | 2855 |
| | | 384 | 44120 | 254.209 | 0.068 | 3738 |
| Delta Wing | 106 564 608 | 304 | 81599 | 783.882 | 0.084 | 9331 |
| | | 552 | 137527 | 2603.362 | 0.162 | 16030 |
| | | 1296 | 195575 | 4888.930 | 0.275 | 17764 |
| | | 2256 | 347693 | 15858.911 | 0.533 | 29763 |
| | | 4532 | 522453 | 38751.447 | 0.958 | 40431 |
| airfoil 30P30N | 760 320 000 | 4800 | 3063581 | > 3 days | 0.719 | >360000 |
| | | 17920 | 8051644 | > 3 days | 3.835 | |

### 3.2 Result analysis and discussion

Table 2 shows that each type of aircraft shape has a fixed number of grid cells. However, in order to meet the requirement of parallel computing, the grid is expected to be split into subblocks, leading to both total grid nodes number *V* and singularity nodes total number $N_s$ increasing[4]. Table 1 also indicates that the number of singularity nodes will increase dramatically as the grid is repartitioned into more subblocks, even if the total size of original grid (counted by the number of grid cells) does not vary. This means in the large scale CFD parallel computing, there exists a large amount of singularity nodes in the multi-block grid, which will also potentially increase the communication cost across blocks.

One of the direct results of large number of singularity nodes is that the reconstruction of singularity connectivity is time-consuming. The plot of running time, *T*, vs. the number of singularity nodes, *Ns*, for the algorithm 1 is shown in Fig. 3(a)(b), and they approximately have a quadratic relationship, i.e.: $T \propto N_s^2$. Similarly the plot of *T* vs. Ns for the improved algorithm 2 is shown in Fig 3(c)(d), and an approximate relation, $T \propto N_s$, can be observed. These results confirmed our analysis of computational complexity for two algorithms in section 2.2.

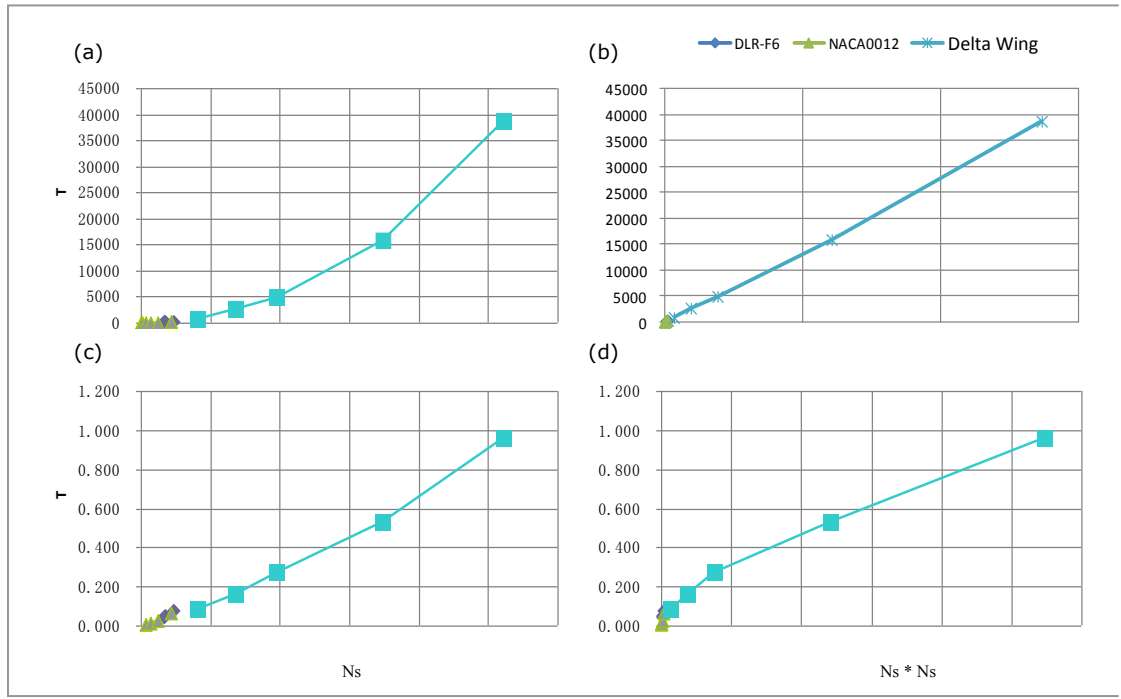

(a) Running time vs. number of singular points in Algorithm 1

(b) Running time vs. the square of number of singular points in Algorithm 1

(c) Running time vs. number of singular points in Algorithm 2

(d) Running time vs. the square of number of singular points in Algorithm 2

Fig. 3 Performance comparison between old algorithm and improved algorithm for reconstructing singularity connectivity. (a) and (b) show the performances of Algorithm 1, while (c) and (d) show the performance of Algorithm 2. In all subplots, the running time (sec.) is shown at the y axis. The number of singular points (denoted by Ns) is shown in subplot (a) and (c), and as a contrast the square of the number of singular points (denoted by Ns*Ns) is shown in subplot (b) and (d).

Measured by wall time, the performance of the improved algorithm 2 is clearly better than the original algorithm 1. Actually the speedup of improved algorithm can reach 2000 or more for a medium-sized case. Take CFD simulation for the case of `Delta Wing` with 106 million grid cells as an example, the original grid is split into 4532 blocks, using our grid repartition tool `TH-MeshSplit`[5,6], for running in 4096 processes, and it costs 9.4 hours by old algorithm, while only less than one second by new algorithm.

## 4 Conclusions

Reconstruction of the singularity connectivity for multi-block structured grid is playing an important role in the parallel simulation of large-scale CFD. In this paper, an improved algorithm is proposed for the reconstruction of singularity connectivity from the available pairwise connections. In order to evaluate the performance of proposed algorithm, 12 test cases are selected and the results show that the improved algorithm can achieve 2000× or much acceleration.

## Acknowledgments

We gratefully thank Ms. XIONG Min for her assistance of polishing the manuscript for better readability.

**References**

[1] B. Basu, S. Enger, M. Breuer, and F. Durst. Three-dimensional simulation of flow and thermal field in a czochralski melt using a block-structured finite-volume method. *Journal of Crystal Growth*, 219: 123–143, 2000.

[2] H. A. Grogger. Finite difference approximations of first derivatives for two-dimensional grid singularities. *Journal of Computational Physics*, 217: 642–657, 2006.

[3] H. A. Grogger. Finite difference approximations of first derivatives for three-dimensional grid singularities. *Journal of Computational Physics*, 225: 2377–2397, 2007.

[4] CB Jenssen, O Fjøsne, O Kloster. A block partitioning algorithm for structured multiblock grids. In: C.A. Lin, A. Ecer, J. Peraux, N. Satofuka, P. Fox, ed. Parallel computational Dynamics. 1999, 205-212

[5] W. Yong-Xian, Z. Li-Lun, L. Wei, C. Yong-Gang, X. Chuan-Fu, W. Zheng-Hua, and Z. Yu. Efficient parallel implementation of large scale 3D structured grid CFD applications on the Tianhe-1A supercomputer. *Computers & Fluids*, 80: 244-250, 2013.

[6] W. Yong-Xian, Z. Li-Lun, C. Yong-Gang, L. Wei, X. Chuan-Fu, and L. Hua-Yong. TH-MeshSplit: a multiblock grid repartitioning tool for parallel CFD applications on heterogeneous CPU/GPU supercomputer. In *24th International Conference on Parallel Computational Fluid Dynamics (ParCFD2012)*, May 21-25, 2012, Atlanta, USA.